\documentclass[twocolumn,prb,nobibnotes,superscriptaddress,floatfix]{revtex4-1} 

\usepackage[dvipdfmx]{graphicx}
\usepackage{amsmath,amssymb,bm}
\usepackage{xcolor}
\usepackage{float}
\usepackage{ulem}

\usepackage[
colorlinks=true,
citecolor=blue,
setpagesize=false]{hyperref}

\begin{document}

\title{
The field-angle anisotropy of proximate Kitaev systems under an in-plane magnetic field
}

\author {Beom Hyun Kim}
\email{bomisu@kias.re.kr}
\affiliation{Korea Institute for Advanced Study, Seoul 02455, South Korea}

\date{\today} 

\begin{abstract}
We have investigated the field-angle behaviors of magnetic excitations
under an in-plane magnetic field for proximate Kitaev systems.
By employing the exact diagonalization method in conjunction with
the linear spin wave theory,
we have demonstrated that the magnetic excitation gap in the polarized phase
is determined by the magnon excitation at $M$ points and
has a strong anisotropy with respect to the field direction 
in the vicinity of the critical field limit.
The specific heat from this magnon excitation bears
qualitatively the same anisotropic behaviors as expected one for
the non-Abelian spin liquid phase in the Kitaev model and
experimentally observed one of the intermediate phases in $\alpha$-RuCl$_3$.
\end{abstract}

\maketitle

\section{Introduction}

\begin{figure}[!b]
\centering
\includegraphics[width=\columnwidth]{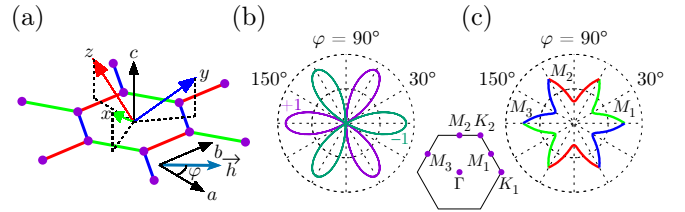}
\caption {
 (a) The connection of three types of bond-dependent Ising interactions,
 and coordinate axes of local spins ($x$, $y$, $z$) and
 global honeycomb lattice ($a$, $b$, $c$) in the Kitaev model.
 The Ising directions in the $x$-, $y$-, and $z$-type bonds drawn
 with green, blue, and red lines are parallel to the $x$, $y$, and $z$
 coordinate axes, respectively.
 The anticipated field-angle variations of (b) Majorana gap of 
 the Kitaev model and (c) magnon gap in the polarized phase of 
 proximate Kitaev systems under an in-plane field.
 $\varphi$ is the angle between the in-plane field and the $a$ axis.
 The radius of curves refers to the size of gap.
 The dark violet (sea green) curve denotes the Chern number $C=+1$ ($-1$).
 When the gap is determined at the $M_1$, $M_2$, and $M_3$ points
 (see the Brillouin zone in inset),
 the curve is drawn with green, red, and blue colors in (c), respectively.
}
\label{fig_ks}
\end{figure}

Quantum fluctuation in frustrated spin systems can prevent
any classical magnetic orders and induce
exotic quantum phases such as a quantum spin liquid (QSL).
The Kitaev model, the ideal $S=\frac{1}{2}$ quantum spin system with
bond-directional Ising interactions in a honeycomb lattice 
(see Fig.~\ref{fig_ks}(a)), is one of intriguing systems
which host the QSL phase as a ground state.
In this exactly solvable model, spin dynamics can be interpreted as
free Majorana fermions in a static $Z_2$ flux~\cite{Kitaev2006}.
Majorana fermions with a gapless energy spectrum leads to the ground QSL phase
and the fractionalized magnetic excitation.

Majorana fermions in the Kitaev model acquire a mass gap $\Delta$ under the magnetic field.
The gapped spectrum stabilizes the topological non-Abelian spin liquid (NASL) phase 
characterized by the Chern number of $C=\pm1$ and protected chiral edge modes~\cite{Kitaev2006,Kasahara2018}.
Because the mass gap is proportional to the multiplication of 
three components of the magnetic field with respect to local spin coordinate axes,
the sign of $C$ and $\Delta$ shows strong field-angle dependency.
For the in-plane field,
the sign change and gap closing occur for the field
applied to the bond direction between nearest neighboring (NN) spins 
as shown in Fig.~\ref{fig_ks}(b)~\cite{Yokoi2021,Hwang2020,Gordon2021}.
The thermal Hall conductivity, specific heat, and magnetotropic coefficient
emulate characteristic features of the Majorana energy spectrum.
The field-angle behaviors of such quantities have been known as
the key feature for the experimental identification of the NALS phase
\cite{Yokoi2021,Hwang2020,Tanaka2020,Gordon2021}.

A lot of theoretical and experimental studies
have been performed to find realistic Kitaev materials~\cite{Winter2017,Takagi2019,Motome2020}.
Among the candidates, $\alpha$-RuCl$_3$ has been turned out to be the best
proximate Kitaev system with dominant Kitaev interaction~\cite{Jackeli2009,HSKim2015}.
The ground state is not the QSL but the antiferromagnetic zigzag order
due to the additional non-Kitaev interactions such as the Heisenberg interactions and 
two types of off-diagonal exchange interactions, called as $\Gamma$ and $\Gamma'$ terms~\cite{Rau2014,Sears2015,Winter2016} (see the general Hamiltonian for the proximate Kitaev system in Eq.~\ref{Eq_SE}). 
Observed magnetic continuum excitations and two-step magnetic entropy release
have been regarded as the evidence of the fractionlized Majorana fermions
\cite{Sandilands2015,Banerjee2016,Yadav2016,Banerjee2017,Do2017,Widmann2019}.
Moreover, recent studies~\cite{Kubota2015,Sears2017,Leahy2017,Baek2017,
Wang2017,Zheng2017,Wolter2017,Winter2018,Banerjee2018,Kasahara2018,Jansa2018,
Wellm2018,Balz2019,Gordon2019,HYLee2020,Yokoi2021,Tanaka2020,Balz2021,Ponomaryov2020,Wulferding2020,BHKim2020} have revealed 
that the zigzag order is easily destroyed when the external magnetic field 
is applied and
the intermediate phase (IP), putative QSL, emerges between the zigzag-order phase and polarized phase.
The NASL phase has been highly anticipated as the IP candidate
because recent thermal Hall conductivity experiments detected
the half-integer plateau and its sign signature under both in-plane
and out-of-plane magnetic fields~\cite{Kasahara2018,Yokoi2021}.
Further, specific heat measurement also probed 
the expected field-angle anisotropy under an in-plane field~\cite{Tanaka2020}.
On the other side, strong sample dependency of the measured quantities
has been reported~\cite{Yamashita2020}. 
A very recent thermal conductivity experiment has reported the emergence of 
different types of QSL phase~\cite{Czajka2021},
and the detailed thermodynamics study~\cite{Bachus2021} has purported
the possibility of the absence of the IP  itself.
All in all, the nature of the IP and/or its existence are still under debate.

In this study, we investigated the field-angle dependence of
magnetic excitation and magnetic specific heat for proximate Kitaev systems
under an in-plane magnetic field.
Using the exact diagonalization (ED) method and linear spin wave theory (LSWT),
we demonstrated that the low-energy excitation features in the polarized phase
for various models relevant to $\alpha$-RuCl$_3$ can be interpreted
in terms of the field-angle anisotropy of magnon gap,
determined at one of three $M$ points depending on the field direction
(see Fig.~\ref{fig_ks}(c)).
The magnetic specific heat dictated by this magnon dynamics shows
qualitatively the same anisotropic behaviors as those in the Kitaev model.
Our result suggests that the anisotropic behaviors in thermodynamic quantities
alone are not a smoking gun of the intermediate NASL phase
in $\alpha$-RuCl$_3$ under the magnetic field, but requires further considerations.

\section{Spin Hamiltonian}

Let $\mathbf{S}_{i}^A$ and $\mathbf{S}_{i}^B$ be two base spins 
at the $i$-th unit cell in a honeycomb lattice.
The general spin Hamiltonian of proximate Kitaev systems with first, second, and
third NN interactions and external magnetic field are given as following:
\begin{align}
H &= \sum_{i\gamma_1} 
\mathbf{S}^A_{i} \cdot \tilde{\mathbf{J}}_{\gamma_1}
\cdot  \mathbf{S}^B_{i_{\gamma_1}} 
+ \sum_{i\gamma_3} 
\mathbf{S}^A_{i} \cdot \tilde{\mathbf{J}}_{\gamma_3}
\cdot  \mathbf{S}^B_{i_{\gamma_3}}  \nonumber \\
&+ \sum_{i\gamma_2} \left(
\mathbf{S}^A_{i} \cdot \tilde{\mathbf{J}}_{\gamma_2}
\cdot  \mathbf{S}^A_{i_{\gamma_2}} 
+
\mathbf{S}^B_{i} \cdot \tilde{\mathbf{J}}_{\gamma_2}
\cdot  \mathbf{S}^B_{i_{\bar{\gamma}_2}} 
\right) \nonumber \\
&- g \mu_B \sum_i \mathbf{h} 
\cdot \left( \mathbf{S}_i^A + \mathbf{S}_i^B \right),
\label{Eq_SE}
\end{align}
where $\gamma_1$, $\gamma_2$, and $\gamma_3$ represent the bond types of 
first, second, and third NN spins, respectively.
$i$ and $i_{\gamma_n}$ are the unit cell indices of two spins in the $\gamma_n$ bond (see Fig.~\ref{fig_kgj}(a)).
$\gamma$ ($=x,y,z$) can be characterized by three coordinate axes of spins.
$\bar{\gamma}_2$ refers to the bond with the opposite direction of the $\gamma_2$ bond.
$g$ is the $g$ factor of spins and $\mu_B$ is the Bohr magneton.
$\tilde{\mathbf{J}}_{\gamma_n}$ is the superexchange dyadic tensor 
of the $\gamma$-type $n$-th NN bond defined as
\begin{align}
\tilde{\mathbf{J}}_{\gamma_n} &=
J_n \hat{\alpha}\hat{\alpha} + J_n \hat{\beta}\hat{\beta} +
(J_n+K_n) \hat{\gamma}\hat{\gamma}
+ \Gamma_n \left(\hat{\alpha}\hat{\beta}+\hat{\beta}\hat{\alpha}\right)
 \nonumber \\
&+ \Gamma'_n \left(
\hat{\alpha}\hat{\gamma}+\hat{\gamma}\hat{\alpha}
+\hat{\gamma}\hat{\beta}+\hat{\beta}\hat{\gamma}
\right),
\label{Eq_Jn}
\end{align}
where $\alpha$, $\beta$, and $\gamma$ are cyclically ordered coordinate axes
of local spins and $\hat{\gamma}$ is the unit vector along the $\gamma$ axis.
$J_n$, $K_n$, $\Gamma_n$, and $\Gamma'_n$ are the exchange parameters 
of the Heisenberg interaction, Kitaev interaction, 
and two types of off-diagonal interactions between $n$-th NN spins, respectively.
The global coordinate axes $a$, $b$, and $c$ can be determined
so that the $a$ ($b$) axis is parallel (perpendicular) to 
the displacement between two spins in the $z_1$-type bond
and the $c$ axis is normal to the honeycomb lattice (See Fig.~\ref{fig_ks}(a)).
Unit vectors $\hat{a}$, $\hat{b}$, and $\hat{c}$ are given as
$\hat{a} =\frac{\hat{x}+\hat{y}-2\hat{z}}{\sqrt{6}}$,
$\hat{b} =\frac{-\hat{x}+\hat{y}}{\sqrt{2}}$,
and $\hat{c} =\frac{\hat{x}+\hat{y}+\hat{z}}{\sqrt{3}}$ 
in terms of local spin coordinate axes.
In the text, the exchange parameters between first NN spins
are simply denoted as $J$, $K$, $\Gamma$, and $\Gamma'$ 
omitting neighbor indices.

\begin{table}[bt]
\caption{Some magnetic models have been proposed for proximate Kitaev system
 $\alpha$-RuCl$_3$ before. The unit of magnetic exchange interactions is meV.}
\label{tb_SE}
\begin{ruledtabular}
\begin{tabular}{c c c c c c c c}
  Model & $J$ & $K$ & $\Gamma$ & $\Gamma'$ & $K_2$ & $J_3$ & Ref. \\
\hline
 1 & $-4.6$ & $ 7.0$ &       &    &   &       & [\onlinecite{Banerjee2016}] \\
 2 & $-1.53$ & $-6.55$ & $5.25$  & $-0.95$  &    &    & [\onlinecite{HSKim2015}] \\
 3 & $-2.7$ & $-10$  & $10.6$ & $-0.9$   &   &    & [\onlinecite{Sears2020}] \\
 4 & $-0.5$ & $-5.0$ & $2.5$ &    &   &  0.5  & [\onlinecite{Winter2018}] \\
 5 & $-1.5$ & $-10$  & $8.8$ &    &   &  0.4  & [\onlinecite{Sears2020}] \\
 6 & $-1.3$ & $-15.1$ & $10.1$ & $-0.1175$  & $-0.68$  & $0.9$  & [\onlinecite{Laurell2020}] \\
 7 & $-4.0$ & $-10.8$ & $5.2$ & $2.9$  &        & $3.26$  & [\onlinecite{Maksimov2020}] \\
\end{tabular}
\end{ruledtabular}
\end{table}

Various magnetic models have been proposed to describe 
the magnetic and thermal properties of proximate Kitaev system $\alpha$-RuCl$_3$
by setting the Kitaev interaction between first NN spins to be dominant and 
specific parameters to be zero in Eq.~\ref{Eq_Jn}~\cite{Yadav2016,Winter2016,Winter2017,HSKim2015,Banerjee2016,Ran2017,Gordon2019,Laurell2020,Hou2017,Eichstaedt2019,Sears2020,Laurell2020,Maksimov2020,Suzuki2021}.
A few representative models, which have been proposed for $\alpha$-RuCl$_3$ before,
are presented in Table~\ref{tb_SE}.

To investigate the field-angle anisotropy of proximate Kitaev model 
under an in-plane magnetic field, 
we first adopted the simple $K$-$\Gamma$-$J_3$ model with $J_3/|K|=\Gamma/|K|=0.1$ and $K<0$, which also accounts for the magnetic phase of the system well.
Then we extended our study to more realistic models presented in Table~\ref{tb_SE}.

\section{ED calculation}

\begin{figure*}[t]
\centering
\includegraphics[width= 2 \columnwidth]{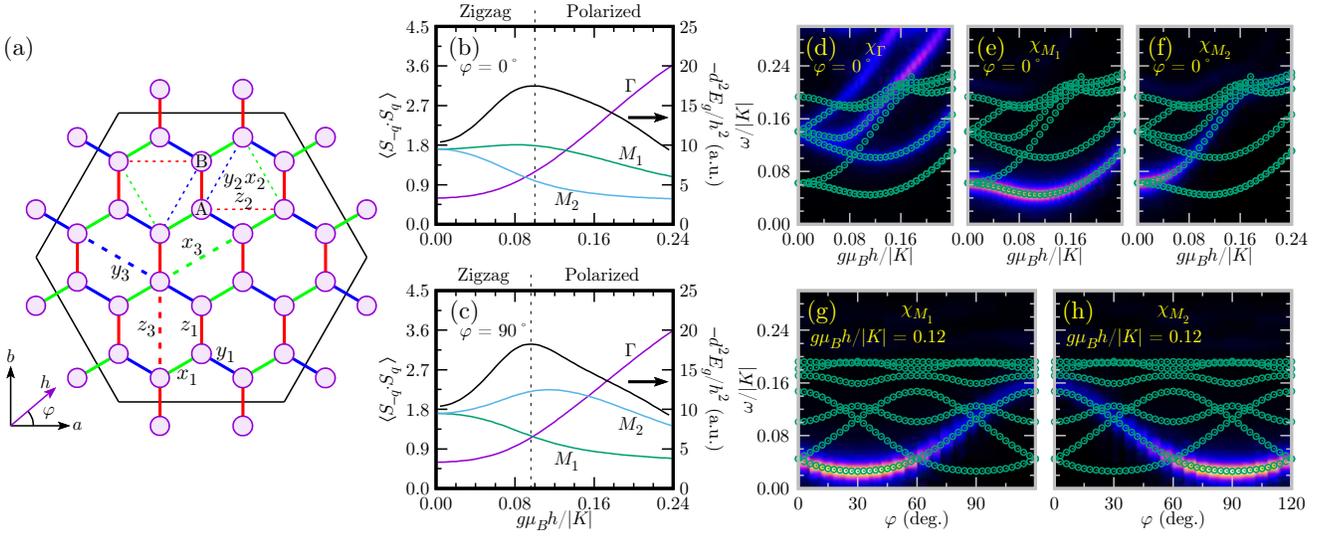}
\caption {
  (a) Schematic diagram of a periodic 24-site cluster.
  `A' and `B' refer to two base sites of a honeycomb lattice, respectively.
  The bonds between first, second, and third nearest neighboring (NN) spins
  are drawn with solid, fine dotted, and thick dotted lines, respectively.
  $x$, $y$, and $z$ types of bonds are highlighted by green, blue, and 
  red colors, respectively.
  $x_1$, $x_2$, and $x_3$ mean the $x$-type first, second, and third
  NN bonds, respectively.
  Static spin structure factors of the ground state
 $\left< \mathbf{S}_{-\mathbf{q}} \cdot \mathbf{S}_{\mathbf{q}} \right>$
  at $\mathbf{q}=\Gamma$, $M_1$, and $M_2$ points under the (b) $a$- 
  and (c) $b$-axis fields ($\varphi=0^\circ$ and $\varphi=90^\circ$).
  Solid black curves in (b) and (c) represent
  the second derivatives of the ground state energy ($-d^2 E_g/d h^2$) 
  with respect to the field strength for given field direction.
  The critical field strength of the phase transition from the zigzag order
  to the polarized phase are indicated with vertical dotted line.
  Dynamical spin structure factors (DSSFs) $\chi_{\mathbf{q}}(\omega)$
  of the $K$-$\Gamma$-$J_3$ model with $J_3/|K|=\Gamma/|K|=0.1$ and $K<0$ at
  (d) $\Gamma$ (e) $M_1$, and (f) $M_2$ points as a function of
  the field strength $h$ when a magnetic field is along the $a$ axis.
  The $\Gamma$, $K_1$, $M_1$, and $M_2$ points are defined in Fig.~\ref{fig_ks}.
  DSSF at (g) $M_1$ and (h) $M_2$ points as a function of the field angle $\varphi$,
  the angle between the in-plane field and the $a$ axis,
  when $g\mu_Bh/|K|= 0.12$.
  All results are calculated by the ED method with a periodic 24-site cluster.
  The circle data represent the lowest seven excitation energies calculated 
  by the thick-restarted Lanczos method~\cite{Wu2000}. 
}
\label{fig_kgj}
\end{figure*}

Employing the ED method with the periodic $24$-site cluster as shown in Fig.~\ref{fig_kgj}(a),
we investigated the magnetic phase transition of the $K$-$\Gamma$-$J_3$ model
with $J_3/|K|=\Gamma/|K|=0.1$ and $K<0$ under an in-plane field.
With the help of the thick-restarted Lanczos method~\cite{Wu2000}, 
we obtained the ground state $\left| \Psi_g \right>$ and its energy $E_g$.
Figure~\ref{fig_kgj}(b) and (c) present the static spin structure factors
(SSSFs)
$\left< \Psi_g \right| \mathbf{S}_{-\mathbf{q}} \cdot \mathbf{S}_{\mathbf{q}} 
\left| \Psi_g \right>$ at $\mathbf{q} = \Gamma$, $M_1$, and $M_2$ points
(see the Brillouin zone in Fig.~\ref{fig_ks})
for the $a$- and $b$-axis fields.
$\mathbf{S}_{\mathbf{q}}$ is defined as 
$\mathbf{S}_\mathbf{q}=\frac{1}{\sqrt{N}}\sum_{j=1}^N{\rm e}^{-i\mathbf{q}\cdot\mathbf{r}_j}\mathbf{S}_j$, where  
$\mathbf{r}_j$ is the position vector of the $j$-th spin $\mathbf{S}_j$ in the honeycomb lattice
and $N$ is the total number of spin sites.
Note that the SSSFs at $\Gamma$ and three $M$ points characterize
the polarized phase and three types of the zigzag order, respectively.
Results indicate that the magnetic phase transition happens
from the zigzag-order phase to polarized phase at around $g\mu_B h/|K| = 0.1$
without hosting any IP under both $a$- and $b$-axis fields.
In the ED calculation, the IP is only feasible for an out-of-plane field.
For the $c$-axis field, the IP appears in the range of 
$0.396 \le g\mu_B h/|K| \le 0.416$ (not shown here).

Further, we numerically explored the magnetic excitation features 
as a function of strength and direction of an in-plane magnetic field
by calculating the dynamical spin structure factor (DSSF)
$\chi_{\mathbf{q}}(\omega)$ as following:
\begin{equation}
\chi_{\mathbf{q}}(\omega)=
-\frac{1}{\pi}\textrm{Im}
\sum_{\nu}
\left< \Psi_g \right|
S_{\mathbf{-q},\nu}\frac{1}{\omega - H + E_g + i\delta}S_{\mathbf{q},\nu} 
\left| \Psi_g \right>,
\end{equation}
where $S_{\mathbf{q},\nu}$ is the $\nu$ ($=x,y,z$) component of $\mathbf{S}_{\mathbf{q}}$
and $\delta$ ($=0.01|K|$) is the broadening parameter. 
When a magnetic field is applied along the $a$ axis
($\varphi=0^\circ$, where $\varphi$ is the angle between the in-plane field and the $a$ axis),
the minimum excitation energies at both the $\Gamma$ and $M_1$ points 
decrease in a weak field limit but
they start to increase at around the critical field corresponding to
the zigzag order to the polarized phase transition.
The excitation gap is determined at the $M_1$ point regardless of the field strength. 
In contrast, the minimum excitation energy at the $M_2$ point,
originally degenerate with those at other $M$ points without the field,
monotonically increases with losing its spectral weight 
when the field strength increases (see Fig.~\ref{fig_kgj}(d), (e), and (f)).

The $24$-site cluster has the dihedral $D_3$ symmetry which includes
both $C_2$ rotation along three NN bond directions and 
$C_3$ rotation along the $c$ axis.
Due to the $C_2$ rotation symmetry, the excitation spectra at three $M$ points
have the $180^\circ$ periodicity for the field angle $\varphi$.
In addition, the excitation spectra at three $M$ points 
have a cyclic relation under the $C_3$ rotation.
These symmetric features are well captured in the polarized phase region
of $g\mu_Bh/|K| = 0.12$ as shown in Fig.~\ref{fig_kgj}(g) and (h).
The DSSFs $\chi_{M_1}$, $\chi_{M_2}$, and $\chi_{M_3}$ have the $180^\circ$ periodicity,
and the excitation gap is determined from $\chi_{M_1}$, $\chi_{M_2}$, and $\chi_{M_3}$
when the field angle $\varphi$ is located 
at $0^\circ \le \varphi \le 60^\circ$, $60^\circ \le \varphi \le 120^\circ$,
and $120^\circ \le \varphi \le 180^\circ$, respectively.
The minima of the excitation gap appear 
whenever a magnetic field is parallel to one out of three NN bond directions
($\varphi = (2n+1) \times 30^\circ$ where $n$ is the integer number).

As shown in Fig.~\ref{fig_kgj}(g) and (h),
the excitation gap at $\varphi=0^\circ$ is higher than that at $\varphi=90^\circ$,
which means that the magnetic entropy under the $b$-axis field ($\varphi=90^\circ$)
is easier to be thermally populated in a low-temperature limit than the $a$-axis-field case ($\varphi=0^\circ$).
Hence, the magnetic specific heat $C_m$ has a lower gap for the field
along the $b$ axis than the $a$ axis, which is well captured in Fig.~\ref{fig_cv}(a).
Here we employed the $K$-$\Gamma$-$J_3$ model with $g\mu_Bh/|K|=0.12$
(see Appendix~\ref{appen:FTLM} for the calculation detail).
Also, we found the six-fold symmetricity of $C_m$ upon the field angle $\varphi$
for the low-temperature case ($k_B T/|K|\lesssim 0.01$) as shown in Fig.~\ref{fig_cv}(b).
$C_m$ periodically varies with minimum values at $\varphi = n\times 60^\circ$
and maximum values at $\varphi = (2n+1)\times 30^\circ$.
The anisotropic behaviors are suppressed when the field strength is
increased from $g\mu_Bh/|K| = 0.12$ to $0.2$, 
which is consistent with the experimental observations of $\alpha$-RuCl$_3$~\cite{Tanaka2020}.

\begin{figure}[t]
\centering
\includegraphics[width= \columnwidth]{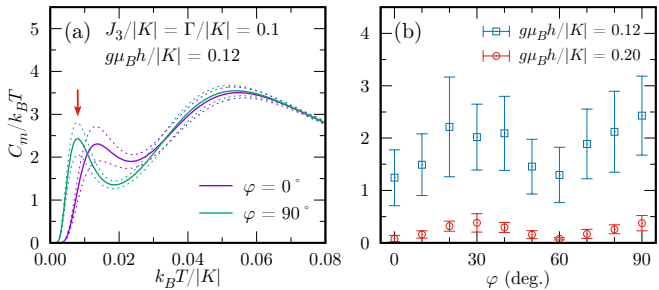}
\caption {
 (a) Theoretical specific heat $C_m$ of the $K$-$\Gamma$-$J_3$ model
 with $J_3/|K|=\Gamma/|K|=0.1$ and $K<0$
 when $g\mu_B h /|K|=0.12$ under the $a$- and $b$-axis fields
 ($\varphi =0^\circ$ and $\varphi = 90^\circ$). 
 (b) $\varphi$ variation of the specific heat
 for $g\mu_B h/|K| = 0.12$ and $0.2$ at $k_B T/|K| =0.008$
 indicated by red arrow in (a).
 Results are calculated by the finite temperature Lanczos method (FTLM)
 \cite{Jaklic2000,Aichhorn2003}
 with a periodic 24-site cluster.
 Dotted lines in (a) and error bars in (b) represent standard deviations of 
 FTLM calculation  (see Appendix~\ref{appen:FTLM} for the detail).
}
\label{fig_cv}
\end{figure}

\section{Spin wave theory}

To get further insight on the field-angle anisotropy under an in-plane magnetic field,
we examined the excitation features in terms of the LSWT (see the detail
in Appendix~\ref{appen:LSWT}).
We performed the LSWT calculation within the polarized phase
in which all spins are ferromagnetically ordered along the field direction.
By reducing the field strength from the infinity, 
we traced the variation of magnon dispersions in the $K$-$\Gamma$-$J_3$ model.

\begin{figure}[t]
\centering
\includegraphics[width=\columnwidth]{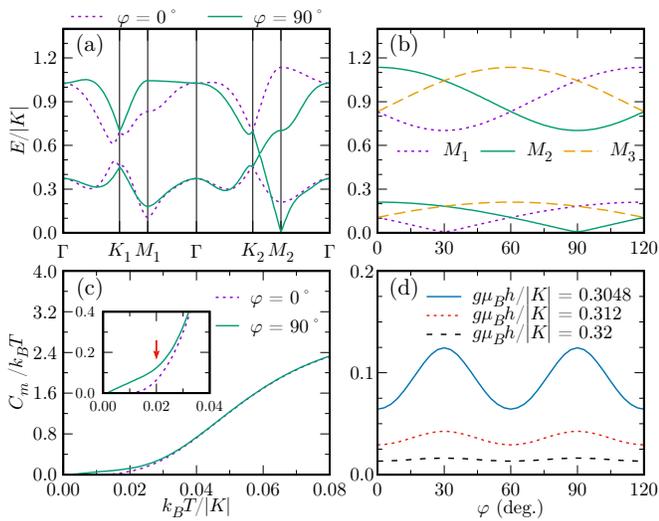}
\caption {
 Magnon bands and magnon specific heats of the $K$-$\Gamma$-$J_3$ model 
 with $J_3/|K|=0.1$, $\Gamma/|K|=0.1$, and $K<0$ when $g\mu_Bh/|K|=0.3048$.
 (a) Magnon bands under the $a$- and $b$-axis fields
 ($\varphi =0^\circ$ and $\varphi = 90^\circ$).
 (b) Magnon bands as a function of $\varphi$ at the $M_1$, $M_2$, and $M_3$ points.
 (c) Specific heats as a function of $T$ under the $a$- and $b$-axis fields.
 Zoom-in of the low-temperature specific heat is presented in the inset of (c).
 (d) Specific heats as a function of $\varphi$ when $k_BT/|K|=0.02$ indicated
 by red arrow in the inset of (c).
 All data are calculated by the linear spin-wave theory.
}
\label{fig_mcv}
\end{figure}

In the classical magnetic phase diagram for $J_3/|K|=\Gamma/|K|=0.1$,
the polarized phase is stabilized when $g\mu_B h/|K|$ is larger than 
about $0.295$ ($0.305$) for the $a$-axis ($b$-axis) field as shown in
the Fig.~\ref{fig_cl} (see the calculation detail in Appendix~\ref{appen:CL}).
The calculated magnon bands are always gapped in this field limit.
The gap is monotonically diminished as the field strength is reduced.
Eventually, the magnon bands condense at the $M_{2,3}$ points ($M_1$ point)
under the $a$-axis ($b$-axis) field with the critical field strength 
of $g\mu_B h/|K| = 0.2930$ ($0.3048$)
(see Fig.~\ref{fig_cl} and Fig.~\ref{fig_mcv}(a)).
For the field-angle dependency, as analogous to the previous DSSFs,
the magnon bands have the $D_3$ character in the critical field limit: 
the low-lying excitation gap at $0^\circ \le \varphi \le 60^\circ$, 
$60^\circ \le \varphi \le 120^\circ$, and $120^\circ \le \varphi \le 180^\circ$,
is determined at $M_1$, $M_2$, and $M_3$ points, respectively.
The gap minima are located at $\varphi=(2n+1) \times 30^\circ$
as shown in Fig.~\ref{fig_mcv}(b).

\begin{figure}[t]
\centering
\includegraphics[width= \columnwidth]{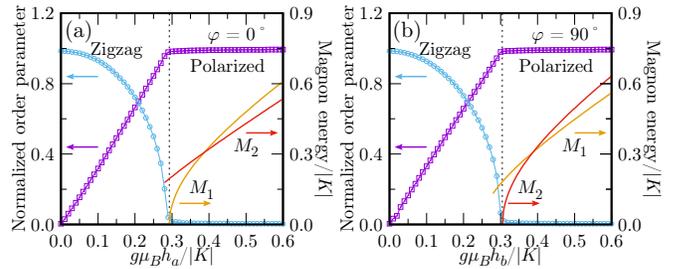}
\caption {
 Expectation values of the order parameters of the zigzag order and
 polarized phase of the classical $K$-$\Gamma$-$J_3$ model with
 $\Gamma/|K|=J_3/|K|=0.1$ and $K<0$ under the (a) $a$- and (b) $b$-aixs fields
 ($\varphi=0^\circ$ and $\varphi=90^\circ$).
 They are calculated with the classical Monte Carlo method at $k_B T/|K|=0.02$.
 The lowest magnon energies at the $M_1$ and $M_2$ points in the polarized phase
 are presented with orange and red lines, respectively.
 Vertical dotted lines represent the phase boundary.
}
\label{fig_cl}
\end{figure}

Based on the magnon dispersions, we calculated the magnon specific heat $C_m$ 
which is attributed to one-magnon excitation.
Figure~\ref{fig_mcv}(c) presents $C_m$ behaviors as a function of temperature $T$
under both $a$- and $b$-axis fields at the critical field strength of $g\mu_B h/|K|=0.3048$.
For the $b$-axis field, 
the $C_m$ is gapless since the magnon is gappless at the critical strength,
while it is still gapped for the $a$-axis field.
The six-fold symmetric behaviors of $C_m$, at low $T$, can be also observed
such that minimum and maximum values at $\varphi = n \times 60^\circ$ and
$(2n+1)\times 30^\circ$ as shown in Fig.~\ref{fig_mcv}(d).
The anisotropic behaviors are progressively suppressed 
as the field strength is increased beyond the critical value.
All behaviors are consistent with both results of ED calculation
and experimental observations of $\alpha$-RuCl$_3$~\cite{Tanaka2020}.

\section{Application to $\alpha$-RuCl$_3$}
\label{Sec:Ru}

\begin{figure}[t]
\centering
\includegraphics[width= \columnwidth]{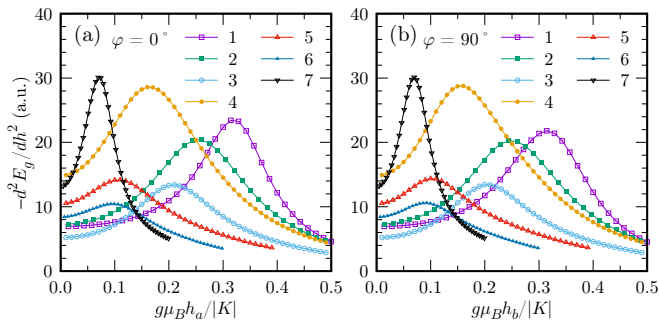}
\caption {
 Second derivative of the ground state energy with respect to 
 the field strength ($-d^2 E_g/dh^2$) for models 1 to 7 in Table~\ref{tb_SE}
 under (a) $a$- and (b) $b$-axis fields ($\varphi=0^\circ$ and $\varphi=90^\circ$).
 The ground state energy is obtained with the ED calculation
 of the periodic 24-site cluster shown in Fig.~\ref{fig_kgj}(a).
}
\label{fig_dea}
\end{figure}

We extended our study to a few representative models (models 1 to 7),
which have been proposed for the magnetic and thermal properties of $\alpha$-RuCl$_3$,
as shown in Table~\ref{tb_SE}.

We investigated their magnetic phase transition under an in-plane magnetic field.
As shown in Fig.~\ref{fig_dea}, the second derivative of their ground state energy
with respect to the field strength ($-d^2 E_g/dh^2$) shows single peak
in the ED calculation.
All models seem to exhibit the phase transition 
from the zigzag order to the polarized phase without hosting any IP
under both $a$- and $b$-axis fields like the $K$-$\Gamma$-$J_3$ model.
However, the ED calculation with small-size cluster is limited
due to the finite size effect.
The absence of IP under an in-plane field is still questionable.

\begin{figure*}[t]
\centering
\includegraphics[width=1.8\columnwidth]{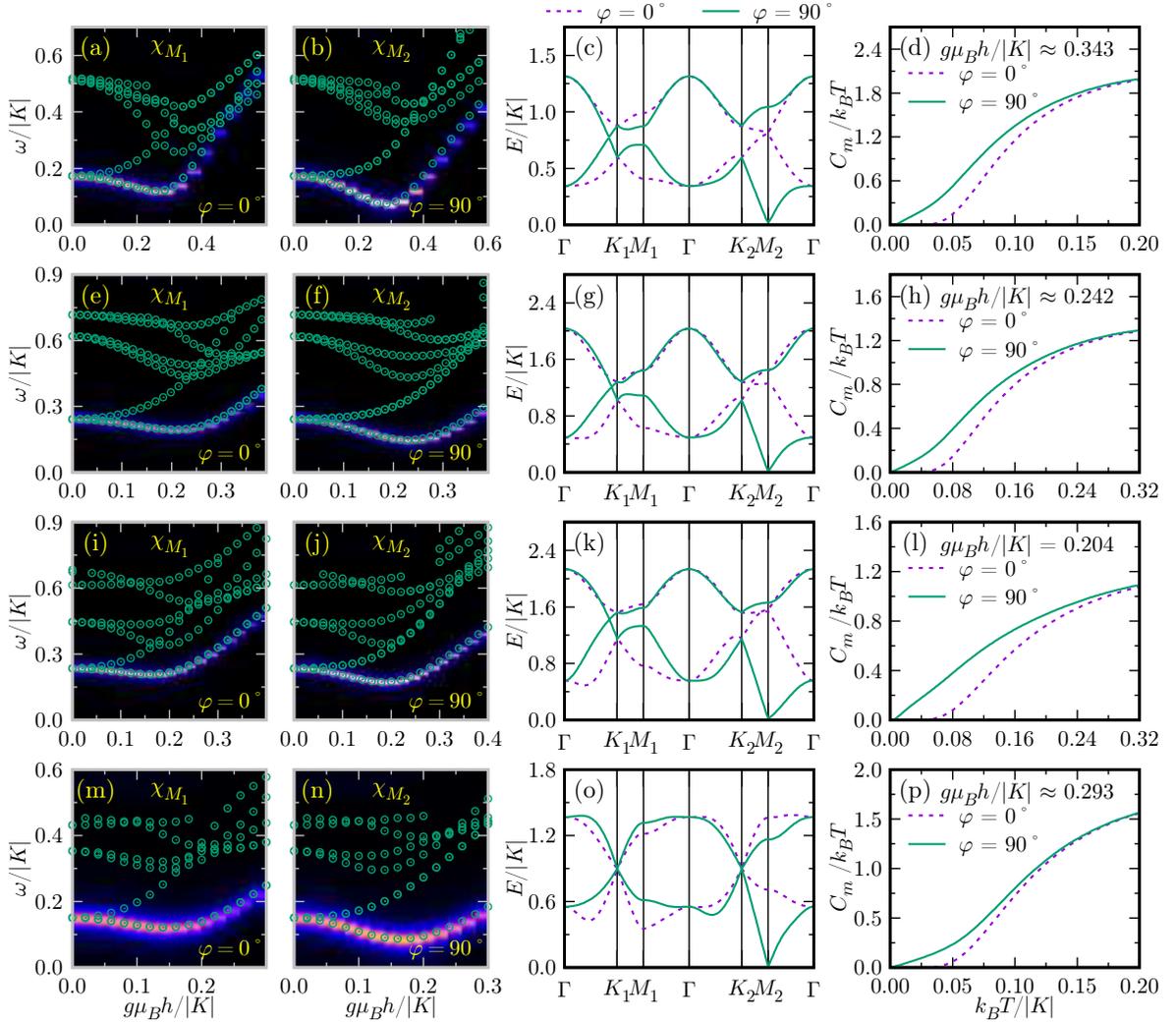}
\caption {
 Dynamical spin structure factors
  $\chi_{M_1}(\omega)$ under the $a$-axis field ($\varphi=0^\circ$)
 and $\chi_{M_2}(\omega)$ under the $b$-axis field ($\varphi=90^\circ$),
 and the magnon dispersions and magnon specific heats
 under the $a$- and $b$-axis fields at the critical field strength for
 (a) -- (d) model 1, (e) -- (h) model 2, (i) -- (l) model 3,
 and (m) -- (p) model 4  in Table~\ref{tb_SE}.
 The critical field strength, at which the magnon bands condense
 under the $b$-axis field, is $g\mu_Bh/|K|\approx 0.343$ for model 1,
 $0.242$ for model 2, $0.204$ for model 3, and $0.293$ for model 4.
 Circle data in (a), (b), (e), (f), (i), (j), (m), and (p) represent
 the lowest seven excitation energies calculated by the thick-restarted 
 Lanczos method~\cite{Wu2000}.
}
\label{fig_mds}
\end{figure*}

\begin{figure*}[t]
\centering
\includegraphics[width=1.8\columnwidth]{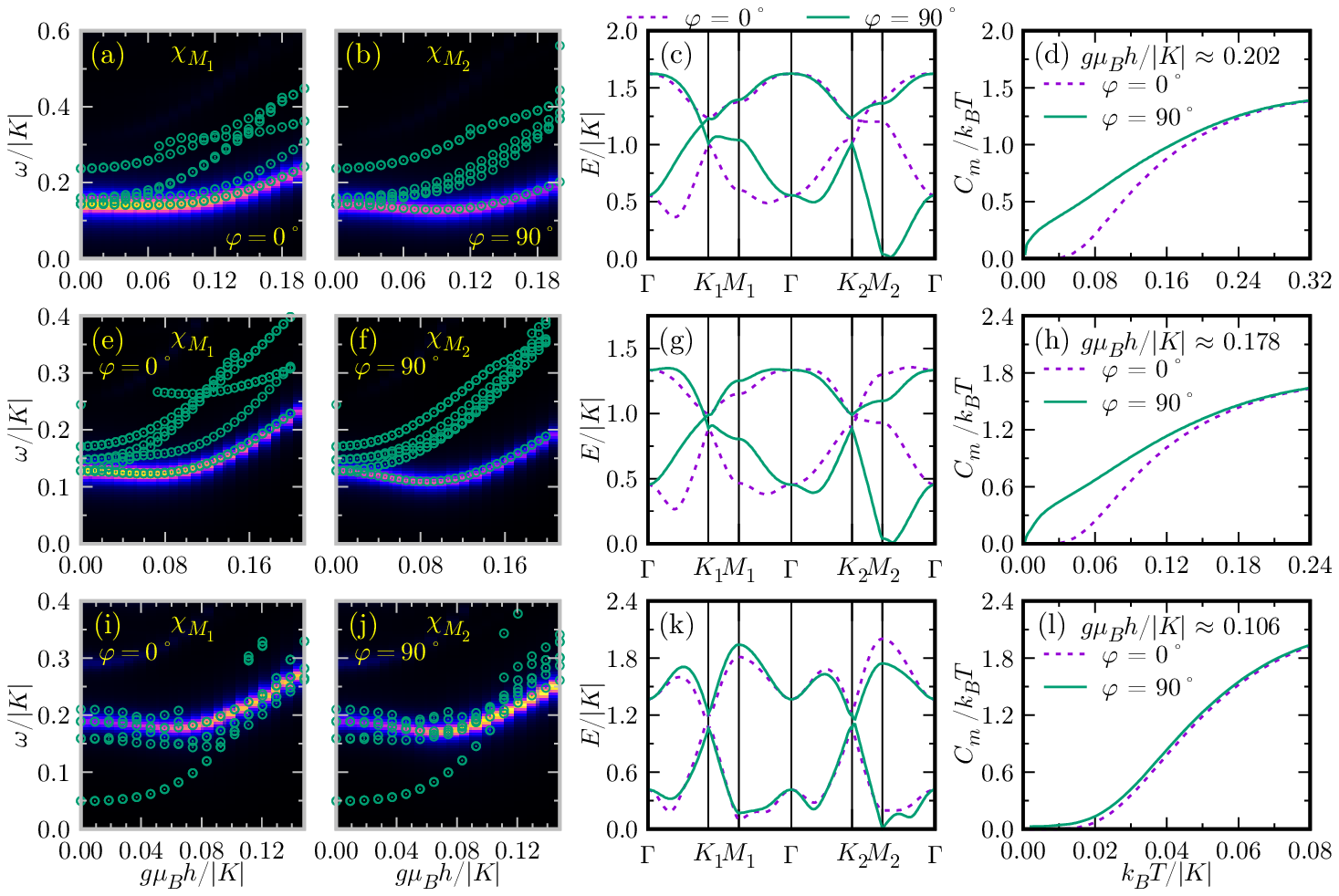}
\caption {
 Dynamical spin structure factors
  $\chi_{M_1}(\omega)$ under the $a$-axis field ($\varphi=0^\circ$)
 and $\chi_{M_2}(\omega)$ under the $b$-axis field ($\varphi=90^\circ$),
 and the magnon dispersions and magnon specific heats
 under the $a$- and $b$-axis fields at the critical field strength for
 (a) -- (d) model 5, (e) -- (h) model 6, and (i) -- (l) model 7 
 in Table~\ref{tb_SE}.
 The critical field strength, at which the magnon bands condense
 under the $b$-axis field, is $g\mu_Bh/|K| \approx 0.202$ for model 5,
 $0.178$ for model 6, and $0.106$ for model 7.
 Circle data in (a), (b), (e), (f), (i), and (j) represent
 the lowest seven excitation energies calculated by the thick-restarted 
 Lanczos method~\cite{Wu2000}.
}
\label{fig_mdsp}
\end{figure*}

We examined the behaviors of DSSFs, magnon dispersions, and magnon specific heat
under both $a$- and $b$-axis fields for models 1--7.
Results are presented in Fig.~\ref{fig_mds} and \ref{fig_mdsp}.
Overall, we found all considered models bear the essentially same field-angle anisotropy.
More specifically, the DSSF, magnon condensation, and magnon specific heat behaviors 
under an in-plane magnetic field are the same between models 1--4 
and the $K$-$\Gamma$-$J_3$ model (see Fig.~\ref{fig_mds}).
Small differences in DSSFs and/or magnon condensation are found for models 5--7:
(i) DSSFs $\chi_{M_1}$ and $\chi_{M_2}$ do not determine 
the excitation gap for $a$- and $b$-axis fields for model 7.
(ii) The magnon condensation point is slightly shifted from the $M_2$ to $\Gamma$ points 
under the $b$-field for models 5 and 6.
Still, the quantitative features remain very similar (see Fig.~\ref{fig_mdsp}).
Therefore, our results suggest the potential role of magnon dynamics 
in explaining the field-angle anisotropy of low-temperature thermal properties for $\alpha$-RuCl$_3$.

Previously, we proposed the $K$-$J_3$ model with $K<0$ and $J_3>0$ 
as the simplest model to account for the IP for the $a$-axis field~\cite{BHKim2020}, where,
we found the magnon simultaneously condenses at both the $M_1$ and $M_2$ points
with the suppression of the anisotropy in the magnon specific heat $C_m(T)$
(see Appendix~\ref{appen:KJ3}).
Hence, we conjecture that the $\Gamma$-term plays a significant role in 
the in-plane anisotropic feature in the proximate Kitaev model with
ferromagnetic $K$ ($K<0$).

\section{Discussion}

In contrast with the ED calculations of various models to show no IP,
plausible evidence of intermediate NASL phase 
such as the half plateau thermal Hall conductivity and 
field-angle anisotropy of specific heat has been reported in $\alpha$-RuCl$_3$
in the intermediate range of the $a$-axis field ($7 \sim 10$ T)~\cite{Yokoi2021,Tanaka2020}.
The lower bound field is coincident with the critical field at which
the zigzag order totally disappears.
However, the upper phase boundary is not evident.
The thermodynamic quantities such as magnetic susceptibility, specific heat,
and magnetic Gr\"{u}neisen parameter show no clear anomalies at around
the upper field~\cite{Balz2019,Balz2021,Bachus2021}.
Moreover, the recent electron spin resonance (ESR) experiment unveiled that
the single-magnon excitation is present across the upper field~\cite{Ponomaryov2020}.
The magnon dynamics of polarized phase is certainly anticipated to emerge 
even for the intermediate region of $\alpha$-RuCl$_3$.

An important remark on the magnon dynamics of the polarized phase
is that the excitation gap is determined at not the $\Gamma$ point but $M$ points
in contrast with the NASL phase in which the magnetic excitation gap
appears at the $\Gamma$ point.
The momentum of the excitation gap could be good measure 
the origin of the field-angle anisotropy of specific heat.
The full magnetic excitation features, which
can be measured by the inelastic neutron scattering (INS) and
resonant inelastic x-ray scattering experiments,
are crucial for determining the nature/existence of IP of $\alpha$-RuCl$_3$.
According to recent INS, Raman spectroscopy, and ESR experiments~\cite{Banerjee2017,Wulferding2020,Ponomaryov2017,Ponomaryov2020},
the minimum excitation energy at the $\Gamma$ point evidently decreases 
before the phase boundary and increases again after the boundary.
In the experimental resolution, however, it is hardly resolved
whether the gap is genuinely closed or not.
It was also verified that the excitation energy at the $M_2$ point in the INS 
is almost constant in the zigzag ordering limit and that it increases
by losing its spectral weight after the phase boundary.
This behavior is quite similar to our ED calculation.
However, only magnetic excitation along the $\Gamma$-$M_2$ line
under the $a$-axis field has been reported in the hitherto INS measurement.
The momentum resolved excitations except for the $\Gamma$ point are limited
in Raman spectroscopy and ESR experiments.
Experimental observations with different momentum lines and 
various field directions are highly required.

The recent theoretical study has pointed out that 
the magnon topology in the polarized phase of the proximate Kitaev system
can give rise to the field-angle variation in the thermal Hall conductivity
which has the same sign structure as that in the NASL phase~\cite{Chern2021}.
Our result supports novel significant feature of this magnon dynamics.
Thus, the field-angle anisotropy of both specific heat and 
thermal Hall conductivity cannot be taken as a key evidence for the NASL phase.
For the ultimate identification, one should test more comprehensively 
a few characteristic features which are inherent to Majorana fermion dynamics,
e.g., the continuum excitation at the $\Gamma$ point, 
the half-integer plateau of thermal Hall conductivity under the $a$-axis field, 
and $T^2$-behavior of the specific heat at low temperature under the $b$-axis field.

\section{Conclusion}

Based on the numerical ED calculation and LSWT analysis,
we have explored the field-angle anisotropy of proximate Kitaev systems 
under an in-plane magnetic field.
We have found that the low-lying excitation gap, interpreted as
the magnon excitation in the polarized phase, is determined at
not the $\Gamma$ but $M$ points in the vicinity of the phase boundary 
between the zigzag order and polarized phase.
Also, the excitation gap has the $60^\circ$ periodicity for the field angle 
with its minimum when the field is along the NN bond direction.
The anisotropy of the low-energy magnon gap can reproduce the field-angle
anisotropy of the specific heat, which was considered a hallmark of the NASL phase.
Our results provide a novel insight into determining the nature/existence 
of IP in the field-induced phase transition of $\alpha$-RuCl$_3$.

\begin{acknowledgments}
{\it Acknowledgements ---}
We acknowledge Bongjae Kim, Kyusung Hwang, Eun-Gook Moon, Kwang-Yong Choi,
Tomonori Shirakawa, Seiji Yunoki, and Young-Woo Son for fruitful discussion.
B.H.K. was supported by KIAS Individual Grants (CG068702). 
Numerical computations have been performed with 
the Center for Advanced Computation Linux Cluster System at KIAS.
\end{acknowledgments}

\appendix

\section{Finite-temperature Lanczos method}
\label{appen:FTLM}

We performed the finite-temperature Lanczos method (FTLM)~\cite{Jaklic2000,Aichhorn2003}
to calculate the magnetic specific heat of proximate Kitaev systems with the 24-site cluster.
The specific heat at temperature $T$ is given as
\begin{equation}
C_m(T) = \frac{k_B}{N(k_BT)^2} \left(
\left< H^2 \right> - \left< H \right>^2
\right),
\end{equation}
where $N$ is the total number of spin sites, $H$ is the spin Hamiltonian,
and $k_B$ is the Boltzmann constant, respectively.
The expectation values of $H$ and $H^2$ can be approximately estimated as following:
\begin{equation}
\left< H \right> \approx \frac{N_{st}}{ZN_{sc}}
\sum_{r=1}^{N_{sc}} \sum_{m=0}^{N_L} \varepsilon_{r,m} 
e^{-\beta \varepsilon_{r,m}},
\end{equation}
\begin{equation}
\left< H^2 \right> \approx \frac{N_{st}}{ZN_{sc}}
\sum_{r=1}^{N_{sc}} \sum_{m=0}^{N_L} (\varepsilon_{r,m} )^2 e^{-\beta \varepsilon_{r,m}},
\end{equation}
where $N_{st}$, $N_{sc}$, and $N_{L}$ are the size of Hilbert space,
the number of initial random states, and the number of Lanczos iteration steps, respectively.
$\varepsilon_{r,m}$ refers to the $m$-th eigenvalue calculated by 
the Lanczos method with the $r$-th initial random state.
$Z$ is the partition function which is also approximately calculated as
\begin{equation}
Z \approx  \frac{N_{st}}{N_{sc}}
\sum_{r=1}^{N_{sc}} \sum_{m=0}^{N_L} e^{-\beta \varepsilon_{r,m}} 
\left| \left< \psi_{r,m} \vert \phi_r \right> \right|^2,
\end{equation}
where $\left| \phi_r \right>$ is the normalized $r$-th initial random state
and $\left| \psi_{r,m} \right>$ is the $m$-th eigenstate calculated by
the Lanczos method with the initial state $\left| \phi_r \right>$.
In the calculation, we set $N_{sc}=250$ and $N_L=200$.

\section{Linear spin wave theory}
\label{appen:LSWT}

In the polarized phase, all spins are ferromagnetically ordered 
along the magnetic field direction.
We performed the liner spin wave theory (LSWT) 
assuming that an in-plane magnetic field is applied away from the $a$ axis
with the field angle $\varphi$.
According to the Holstein-Primakoff transformation, the spin operators
can be written in terms of two bosonic operators as following:
\begin{subequations}
\begin{align}
S_{ix'}^A &= \frac{\sqrt{2S-a_i^\dagger a_i} a_i + 
   a_i^\dagger \sqrt{2S-a_i^\dagger a_i}}{2}
   \nonumber \\
   &\approx \sqrt{\frac{S}{2}}\left( a_i + a_i^\dagger  \right),
\end{align}
\begin{align}
S_{iy'}^A &= \frac{\sqrt{2S-a_i^\dagger a_i} a_i - 
   a_i^\dagger \sqrt{2S-a_i^\dagger a_i}}{2i}
   \nonumber \\
   &\approx -i\sqrt{\frac{S}{2}}\left( a_i - a_i^\dagger  \right),
\end{align}
\begin{equation}
S_{iz'}^A = S - a_i^\dagger a_i,
\end{equation}
\begin{equation}
S_{ix'}^B 
   \approx \sqrt{\frac{S}{2}}\left( b_i + b_i^\dagger  \right),
\end{equation}
\begin{equation}
S_{iy'}^B 
   \approx -i\sqrt{\frac{S}{2}}\left( b_i - b_i^\dagger  \right),
\end{equation}
\begin{equation}
S_{iz'}^B = S - b_i^\dagger b_i,
\end{equation}
\end{subequations}
where $a^\dagger_i$ ($a_i$) and $b^\dagger_i$ ($b_i$) are 
the bosonic creation (annihilation) operators of magnons 
at the $i$-th $A$ and $B$ sublattices, respectively, and
$x'$, $y'$, and $z'$ are coordinate axes defined as
\begin{subequations}
\begin{equation}
\hat{x}' =-\hat{c},
\end{equation}
\begin{equation}
\hat{y}' = -\sin \varphi \hat{a} + \cos \varphi \hat{b},
\end{equation}
\begin{equation}
\hat{z}' = \cos \varphi \hat{a} + \sin \varphi \hat{b},
\end{equation}
\end{subequations}
where $a$, $b$, and $c$ are global coordinate axes of the lattice.

The magnetic interactions between first, second, and third NN spins in Eq.~\ref{Eq_SE}
can be approximately expressed in terms of the bosonic operators 
as following:
\begin{multline}
\mathbf{S}^A_{i} \cdot \tilde{\mathbf{J}}_{\gamma_n}
\cdot  \mathbf{S}^B_{i_{\gamma_n}} \approx
Sc_{\gamma_n}(\varphi)
\left( S - a_i^\dagger a_i - b_{i_{\gamma_n}}^\dagger b_{i_{\gamma_n}}
\right) \\
+ Sd_{\gamma_n}(\varphi) a_i^\dagger b_{i_{\gamma_n}}^\dagger
+Sd_{\gamma_n}(\varphi)^* a_i b_{i_{\gamma_n}} \\
+Sh_{\gamma_n}(\varphi) a_i^\dagger b_{i_{\gamma_n}} 
+Sh_{\gamma_n}(\varphi)^* a_i b_{i_{\gamma_n}}^\dagger,
\end{multline}
\begin{multline}
\mathbf{S}^A_{i} \cdot \tilde{\mathbf{J}}_{\gamma_2} (\varphi)
\cdot  \mathbf{S}^A_{i_{\gamma_2}} 
+
\mathbf{S}^B_{i} \cdot \tilde{\mathbf{J}}_{\gamma_2} (\varphi)
\cdot  \mathbf{S}^B_{i_{\bar{\gamma}_2}} 
\\ \approx
Sc_{\gamma_2}(\varphi) \left( 2S - a_i^\dagger a_i 
-a_{i_{\gamma_2}}^\dagger a_{i_{\gamma_2}} -b_i^\dagger b_i 
-b_{i_{\bar{\gamma}_2}}^\dagger b_{i_{\bar{\gamma}_2}}
\right) \\
+Sd_{\gamma_2}(\varphi) a_i^\dagger a_{i_{\gamma_2}}^\dagger
+Sd_{\gamma_2}(\varphi)^* a_i a_{i_{\gamma_2}} \\
+Sh_{\gamma_2}(\varphi) a_i^\dagger a_{i_{\gamma_2}} 
+Sh_{\gamma_2}(\varphi)^* a_i a_{i_{\gamma_2}}^\dagger \\
+Sd_{\gamma_2}(\varphi) b_i^\dagger b_{i_{\bar{\gamma}_2}}^\dagger
+Sd_{\gamma_2}(\varphi)^* b_i b_{i_{\bar{\gamma}_2}} \\
+Sh_{\gamma_2}(\varphi) b_i^\dagger b_{i_{\bar{\gamma}_2}} 
+Sh_{\gamma_2}(\varphi)^* b_i b_{i_{\bar{\gamma}_2}}^\dagger,
\end{multline}
and
\begin{equation}
\mathbf{h} \cdot \left( \mathbf{S}_i^A + \mathbf{S}_i^B \right)
= h \left( 2S - a_i^\dagger a_i - b_i^\dagger b_i \right),
\end{equation}
where 
\begin{subequations}
\begin{equation}
c_{\gamma_n} (\varphi) = \tilde{J}_{\gamma_n}^{z'z'}
\end{equation}
\begin{equation}
h_{\gamma_n}(\varphi) = \frac{1}{2}\left(
  \tilde{J}_{\gamma_n}^{x'x'} + \tilde{J}_{\gamma_n}^{y'y'}
-i \tilde{J}_{\gamma_n}^{x'y'} +i \tilde{J}_{\gamma_n}^{y'x'} \right),
\end{equation}
\begin{equation}
d_{\gamma_n}(\varphi) = \frac{1}{2}\left(
  \tilde{J}_{\gamma_n}^{x'x'} - \tilde{J}_{\gamma_n}^{y'y'}
+i \tilde{J}_{\gamma_n}^{x'y'} +i \tilde{J}_{\gamma_n}^{y'x'} \right).
\end{equation}
\end{subequations}
Here $\tilde{J}_{\gamma_n}^{\alpha\beta}
= \hat{\alpha} \cdot \tilde{\mathbf{J}}_{\gamma_n} \cdot \hat{\beta}$.
We define the spinor operators 
$\bm{\psi}_i^\dagger = \begin{pmatrix} a^\dagger_i & b^\dagger_i \end{pmatrix}$,
$\bm{\psi}_i = \begin{pmatrix} a_i & b_i \end{pmatrix}^\intercal$, and
$\bm{\psi}_i^* = \begin{pmatrix} a^\dagger_i & b^\dagger_i \end{pmatrix}^\intercal$.
Equation~\ref{Eq_SE} can be given as following:
\begin{widetext}
\begin{align}
H & \approx -N_u S^2 E (h,\varphi)
+S\epsilon (h,\varphi) \sum_i \bm{\psi}^\dagger_i \bm{\psi}_i 
+S \sum_{i\gamma_1} \bm{\psi}^\dagger_{i_{\gamma_1}}
\begin{pmatrix}
0 & 0 \\ h_{\gamma_1} (\varphi)^* & 0 
\end{pmatrix} \bm{\psi}_i
+S \sum_{i \gamma_1} \bm{\psi}^\dagger_{i_{\bar{\gamma}_1}}
\begin{pmatrix}
0 & h_{\gamma_1} (\varphi) \\ 0 & 0 
\end{pmatrix} \bm{\psi}_i  \nonumber \\
&+ \frac{S}{2} \sum_{i\gamma_1} 
\left[
\bm{\psi}^\dagger_{i_{\gamma_1}}
\begin{pmatrix}
0 & 0 \\ d_{\gamma_1}(\varphi) & 0
\end{pmatrix} \bm{\psi}_i^*
+h.c. \right]
+\frac{S}{2} \sum_{i\gamma_1} \left[
\bm{\psi}^\dagger_{i_{\bar{\gamma}_1}}
\begin{pmatrix}
0 & d_{\gamma_1}(\varphi)  \\ 0 & 0
\end{pmatrix} \bm{\psi}_i^* +h.c.
\right] \nonumber \\
&+S \sum_{i\gamma_2} \bm{\psi}^\dagger_{i_{\gamma_2}}
\begin{pmatrix}
h_{\gamma_2} (\varphi)^*  & 0 \\ 0 & h_{\gamma_2} (\varphi)
\end{pmatrix} \bm{\psi}_i
+S \sum_{i \gamma_2 } \bm{\psi}^\dagger_{i_{\bar{\gamma}_2}}
\begin{pmatrix}
h_{\gamma_2} (\varphi) & 0 \\ 0 & h_{\gamma_2} (\varphi)^*
\end{pmatrix} \bm{\psi}_i  \nonumber \\
&+\frac{S}{2} \sum_{i{\gamma_2}} 
\left[ \bm{\psi}^\dagger_{i_{\gamma_2}}
\begin{pmatrix}
d_{\gamma_2}(\varphi) & 0 \\ 0 & d_{\gamma_2}(\varphi) 
\end{pmatrix} \bm{\psi}_i^* +h.c. \right]
+\frac{S}{2} \sum_{i\gamma_2} \left[
\bm{\psi}^\dagger_{i_{\bar{\gamma}_2}}
\begin{pmatrix}
d_{\gamma_2}(\varphi) & 0 \\ 0 & d_{\gamma_2}(\varphi) 
\end{pmatrix} \bm{\psi}_i^* + h.c.  \right] \nonumber \\
&+S \sum_{i\gamma_3} \bm{\psi}^\dagger_{i_{\gamma_3}}
\begin{pmatrix}
0 & 0 \\ h_{\gamma_3} (\varphi)^* & 0 
\end{pmatrix} \bm{\psi}_i
+S \sum_{i \gamma_3} \bm{\psi}^\dagger_{i_{\bar{\gamma}_3}}
\begin{pmatrix}
0 & h_{\gamma_3} (\varphi) \\ 0 & 0 
\end{pmatrix} \bm{\psi}_i  \nonumber \\
&+ \frac{S}{2} \sum_{i\gamma_3} \left[
\bm{\psi}^\dagger_{i_{\gamma_3}}
\begin{pmatrix}
0 & 0 \\ d_{\gamma_3}(\varphi) & 0
\end{pmatrix} \bm{\psi}_i^* +h.c. \right]
+\frac{S}{2} \sum_{i\gamma_3} \left[
\bm{\psi}^\dagger_{i_{\bar{\gamma}_3}}
\begin{pmatrix}
0 & d_{\gamma_3}(\varphi)  \\ 0 & 0
\end{pmatrix} \bm{\psi}_i^* +h.c. \right],
\label{Eq_SP}
\end{align}
\end{widetext}
where $\epsilon (h,\varphi) = \frac{g\mu_B h}{S} -
\sum_{\gamma_1} c_{\gamma_1} (\varphi)
-2\sum_{\gamma_2} c_{\gamma_2} (\varphi)
-\sum_{\gamma_3} c_{\gamma_3} (\varphi)$, 
$E (h,\varphi) = \epsilon (h,\varphi) + \frac{g\mu_B h}{S}$,
and $N_u$ is the number of unit cells in the lattice.

To rewrite the Hamiltonian in momentum space, we performed
the Fourier transformation of spinor operators like
$\bm{\psi}^\dagger_{i}=
\frac{1}{\sqrt{N_u}} \sum_{\mathbf{k} \in BZ}
e^{-i\mathbf{k}\cdot \mathbf{r}_i} 
\bm{\psi}_{\mathbf{k}}^\dagger$ and
$\bm{\psi}_{\mathbf{k}}  =
\frac{1}{\sqrt{N_u}} \sum_i e^{ -i\mathbf{k} \cdot \mathbf{r}_i} 
\bm{\psi}_i$ where $\mathbf{r}_i$ is the position vector of
the $i$-th unit cell.
Equation~\ref{Eq_SP} can be transformed as following:
\begin{multline}
H \approx -N_u S^2 E(h,\varphi) \\
+S
\sum_{\mathbf{k}\in BZ}
 \bm{\psi}^\dagger_{\mathbf{k}} 
\begin{pmatrix}
 h_{2} ( h, \varphi, \mathbf{k}) &  h_1 ( \varphi, \mathbf{k})  \\
 h_1 ( \varphi, \mathbf{k})^* & h_{3} ( h, \varphi, \mathbf{k})
\end{pmatrix}
\bm{\psi}_{\mathbf{k}} \\
+\frac{S}{2}
\sum_{\mathbf{k}\in BZ}
\left[
 \bm{\psi}^\dagger_{\mathbf{k}} 
\begin{pmatrix}
d_2 (\varphi, \mathbf{k}) &  d_1 (\varphi, \mathbf{k}) \\
 d_1 (\varphi, -\mathbf{k})  & d_2 (\varphi, \mathbf{k})
\end{pmatrix}
\bm{\psi}_{-\mathbf{k}}^*
+h.c.  \right],
\label{Eq_SK}
\end{multline}
where 
\begin{subequations}
\begin{equation}
h_1 ({\varphi},\mathbf{k}) = 
\sum_{\gamma_1} h_{\gamma_1}(\varphi)
e^{i\mathbf{k}\cdot\mathbf{r}_{\gamma_1}}
+ \sum_{\gamma_3} h_{\gamma_3}(\varphi)
e^{i\mathbf{k}\cdot\mathbf{r}_{\gamma_3}},
\end{equation}
\begin{multline}
h_2 (h,{\varphi},\mathbf{k}) = \epsilon (h,\varphi) \\
 + \sum_{\gamma_2} \left[
 h_{\gamma_2}(\varphi)e^{i\mathbf{k}\cdot\mathbf{r}_{\gamma_2} }
+h_{\gamma_2}(\varphi)^*e^{-i\mathbf{k}\cdot\mathbf{r}_{\gamma_2} }
\right],
\end{multline}
\begin{multline}
h_3 (h,{\varphi},\mathbf{k}) = \epsilon (h,\varphi) \\
 + \sum_{\gamma_2} \left[
 h_{\gamma_2}(\varphi)e^{-i\mathbf{k}\cdot\mathbf{r}_{\gamma_2} }
+h_{\gamma_2}(\varphi)^*e^{i\mathbf{k}\cdot\mathbf{r}_{\gamma_2} }
\right],
\end{multline}
\begin{equation}
d_1 ( \varphi, \mathbf{k}) = 
\sum_{\gamma_1} d_{\gamma_1}(\varphi) 
e^{i\mathbf{k}\cdot \mathbf{r}_{\gamma_1}} +
\sum_{\gamma_3} d_{\gamma_3}(\varphi) 
e^{i\mathbf{k}\cdot \mathbf{r}_{\gamma_3}},
\end{equation}
\begin{equation}
d_2 ( \varphi, \mathbf{k}) = \sum_{\gamma_2} 
\left[
d_{\gamma_2}(\varphi) 
e^{i\mathbf{k}\cdot \mathbf{r}_{\gamma_2}} 
+ d_{\gamma_2}(\varphi) 
e^{i\mathbf{k}\cdot \mathbf{r}_{\bar{\gamma}_2}} 
\right],
\end{equation}
\end{subequations}
where $\mathbf{r}_{\gamma_n} = -\mathbf{r}_{\bar{\gamma}_n} = \mathbf{r}_{i_{\gamma_n}}-\mathbf{r}_i$.
According to the Bogoliubov transformation, Equation~\ref{Eq_SK} is extended 
as following:
\begin{multline}
H  \approx -N_u S\left[ S E (h,\varphi) + \epsilon (h,\varphi) \right] \\
  + \frac{S}{2} \sum_{\mathbf{k}\in BZ}
\begin{pmatrix}
\bm{\psi}_{\mathbf{k}}^\dagger &
\bm{\psi}_{-\mathbf{k}}^\intercal 
\end{pmatrix}
\begin{pmatrix}
\mathbf{h}(h,\varphi,\mathbf{k}) &  \mathbf{d} (\varphi,\mathbf{k}) \\
\mathbf{d}(\varphi,\mathbf{k})^\dagger & \mathbf{h} (h,\varphi,-\mathbf{k})^\intercal
\end{pmatrix}
\begin{pmatrix}
\bm{\psi}_{\mathbf{k}} \\
\bm{\psi}_{-\mathbf{k}}^* 
\end{pmatrix},
\end{multline}
where
\begin{subequations}
\begin{equation}
\mathbf{h}(h, \varphi, \mathbf{k}) =
\begin{pmatrix}
 h_{2} ( h, \varphi, \mathbf{k}) &  h_1 ( \varphi, \mathbf{k})  \\
 h_1 ( \varphi, \mathbf{k})^* & h_{3} ( h, \varphi, \mathbf{k})
\end{pmatrix},
\end{equation} 
\begin{equation}
\mathbf{d}(\varphi, \mathbf{k}) =
\begin{pmatrix}
d_2 (\varphi, \mathbf{k}) &  d_1 (\varphi, \mathbf{k}) \\
 d_1 (\varphi,\mathbf{k})^\dagger  & d_2 (\varphi, \mathbf{k})
\end{pmatrix}.
\end{equation}
\end{subequations}
We obtained the dynamic equation of motion of spinor operators as follows:
\begin{equation}
i\hbar \partial_t 
\begin{pmatrix}
\bm{\psi}_{\mathbf{k}} \\
\bm{\psi}_{-\mathbf{k}}^* 
\end{pmatrix} =
\begin{pmatrix}
\mathbf{h}(h,\varphi,\mathbf{k}) &  \bm{\Delta} (\varphi,\mathbf{k}) \\
-\bm{\Delta}(\varphi,\mathbf{k})^\dagger & -\mathbf{h} (h,\varphi,-\mathbf{k})^\intercal
\end{pmatrix}
\begin{pmatrix}
\bm{\psi}_{\mathbf{k}} \\
\bm{\psi}_{-\mathbf{k}}^* 
\end{pmatrix},
\label{Eq_DM}
\end{equation}
where $\bm{\Delta}(\varphi,\mathbf{k}) = \frac{1}{2}
\left[ \mathbf{d} (\varphi,\mathbf{k})+\mathbf{d} (\varphi,-\mathbf{k})^\intercal \right]$.
By solving the general eigenvalue problem of Eq.~\ref{Eq_DM}, we calculated magnon dispersions.

Let $\epsilon_n({\mathbf{k}})$ be the $n$-th magnon band at a given momentum $\mathbf{k}$ ($n=1,2$).
The average energy per unit cell can be given as
\begin{equation}
e(T) \approx -S^2 E (h,\varphi)+ \frac{S}{N_u}
\sum_{n,\mathbf{k}\in BZ} \frac{\epsilon_n(\mathbf{k})}{e^{\beta \epsilon_n(\mathbf{k})}-1},
\end{equation}
where $\beta=k_B T$ because of the Bose-Einstein statistics of magnons.
The magnon specific heat is deduced as
\begin{equation}
C_m(T) = \frac{d}{dT} e(T) = k_B \frac{S}{N_u} \sum_{n,\mathbf{k}\in BZ} 
\frac{ \beta^2 \epsilon_n(\mathbf{k})^2 e^{\beta \epsilon_n(\mathbf{k})}}
{(e^{\beta \epsilon_n(\mathbf{k})}-1)^2}.
\end{equation}

\section{Classical phase diagram}
\label{appen:CL}

To determine the classical phase diagram of proximate Kitaev systems
under an in-plane field, we performed 
the classical Monte Carlo (MC) calculation with the standard Metropolis algorithm. 
By considering a periodic $2\times 36 \times 36$ cluster,
we ran the 40000 MC steps after 20000 MC steps for thermalization 
at $k_B T/|K|=0.02$
and calculated the expectation value of order parameters for
both the zigzag order and the polarized phase as a function of the field strength.

\section{$K$-$J_3$ model}
\label{appen:KJ3}

\begin{figure}[t]
\centering
\includegraphics[width=\columnwidth]{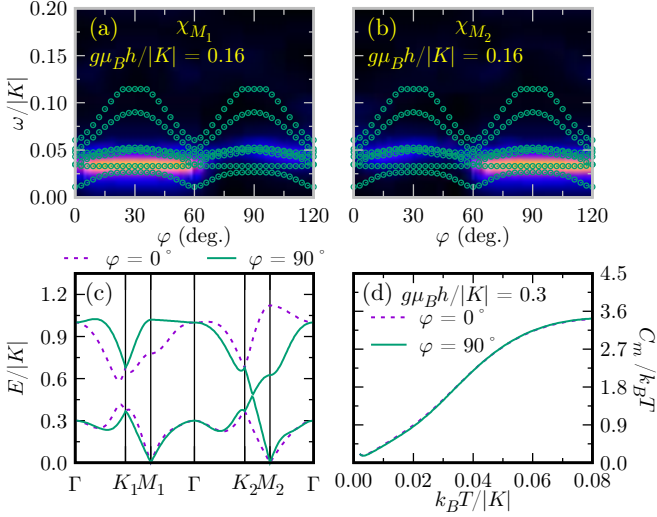}
\caption {
 Dynamical spin structer factors
 (a) $\chi_{M_1}(\omega)$ and (b) $\chi_{M_2}(\omega)$
 as a function of a field angle $\varphi$ in the $K$-$J_3$ model
 with $J_3/|K|=0.1$, $K<0$, and $g\mu_Bh/|K| = 0.16$.
 (c) Magnon dispersions and (d) magnon specific heats
 under both $a$- and $b$-axis fields ($\varphi=0^\circ$ and $\varphi=90^\circ$)
 at the critical field strength 
 ($g\mu_Bh/|K| = 0.15$) in the $K$-$J_3$ model.
 Circle data in (a) and (b) represent the lowest seven excitation energies 
 calculated by the thick-restarted Lanczos method~\cite{Wu2000}.
}
\label{fig_kj3}
\end{figure}

The $K$-$J_3$ model with $K<0$ and $J_3>0$ has been proposed as 
the simplest proximate Kitaev model which shows the genuine intermediate phase 
under both the $a$- and $c$-axis fields~\cite{BHKim2020}.
We have explored the behaviors of DSSFs, magnon dispersions,
and magnon specific heat of the $K$-$J_3$ model under in-plane magnetic fields.
As shown in Fig.~\ref{fig_kj3}, 
the field-angle anisotropy in the $K$-$J_3$ model is less evident 
than those in other models.
The DSSFs $\chi_{M_1}$, $\chi_{M_2}$, and $\chi_{M_3}$ are almost constant
in the range of $0^\circ \le \varphi \le 60^\circ$, 
$60^\circ \le \varphi \le 120^\circ$, and $120^\circ \le \varphi \le 180^\circ$,
respectively.
The magnon bands in the polarized phase condense almost simultaneously
at the three $M$ points and the field-angle anisotropy of magnon specific heat 
is almost suppressed.

\end{document}